\documentclass[11pt,titlepage]{article}

\usepackage{setspace} 

\usepackage{amssymb,amsmath,bm}
\usepackage[colorlinks]{hyperref}

\usepackage{graphicx}

\usepackage[round,numbers,sort&compress]{natbib} 
\title{Quantitative analysis of competition in post-transcriptional regulation reveals a novel signature in target expression variation}

\author{Filippos~D~Klironomos\thanks{
           Corresponding author.  Email: fklirono@uni-koeln.de Address: 
           Institute for Theoretical Physics,
	   University of Cologne,
	   Z\"ulpicher Stra{\ss}e 77, 50937 Cologne, Germany,
	   Tel.:~+49 221 470 2178, Fax:~+49 221 470 5159} \\
	Institute for Theoretical Physics, \\
	University of Cologne, Cologne, Germany
	\and Johannes Berg \\
	Institute for Theoretical Physics, \\
	University of Cologne, Cologne, Germany\\ and \\
    Systems Biology of Ageing - Sybacol,\\
    Cologne, Germany}

\date{}

\begin{document}

\maketitle

\abstract{
When small RNAs are loaded onto Argonaute proteins they can form the RNA-induced silencing complexes (RISCs), which mediate RNA interference (RNAi).
RISC-formation is dependent on a shared pool of Argonaute proteins and RISC-loading factors, and is susceptible to competition among small RNAs.
We present a mathematical model that aims to understand how small RNA competition for RISC-formation affects target gene repression.
We discuss that small RNA activity is limited by RISC-formation, RISC-degradation and the availability of Argonautes.
We show that different competition conditions for RISC-loading result in different signatures of RNAi determined also by the amount of RISC-recycling taking place.
In particular, we find that the small RNAs less efficient at RISC-formation, can perform in the low RISC-recycling range equally well as their more effective counterparts.
Additionally, we predict that under conditions of low RISC-loading efficiency and high RISC-recycling,
the variation in target levels increases linearly with the target transcription rate.
Furthermore, we show that RISC-recycling determines the effect that Argonaute scarcity conditions have on target expression variation.
Our observations taken together offer a framework of predictions which can be used in order to infer from data the particular characteristics of underlying RNAi activity.

\clearpage

\section*{Introduction}

Post-transcriptional regulation (PTR) is mediated by RNA-induced silencing complexes (RISCs) assembled from Argonaute proteins and microRNA (miRNA) or small-interfering RNA (siRNA) 
molecules\citep{Kim2009}.
RISCs act on mRNA transcripts via the RNA-interference pathway increasing the mRNA turnover (degradation) rate via cleaving\citep{Rhoades2006} or destabilization\citep{Guo2010}.
A single RISC can act multiple times like an enzyme\citep{Hutvagner2002}, even when target destabilization rather than cleaving takes place\citep{Haley2004,Wu2006,Baccarini2011}.
mRNA destabilization and cleaving differ in the number of times a RISC can operate before dissociation or degradation of the loaded small RNA.
For example, \textit{let-7} miRNA-programmed RISCs were found to catalyze each approximately 10 target molecules\citep{Hutvagner2002}.
Another study showed that \textit{let-7} siRNA-programmed RISCs participated each in 50 rounds of catalysis on average\citep{Haley2004}.
However, experiments using miR223-programmed RISCs constitutively expressed in human 293T cells reported an average of only two rounds of activity per RISC\citep{Baccarini2011}.
Conceptually, it is simpler to study PTR in the two limits of RISC operational activity\citep{Levine2007,Shimoni2007,Hao2011,Loinger2012}:
the non-enzymatic stoichiometric mode, where each RISC degrades with the transcript it has targeted,
and the fully enzymatic catalytic mode, where RISC-degradation is independent of mRNA-targeting.
Part of the scope of this work is to understand the dynamics of PTR activity in these two distinct operational regimes.

Post-transcriptional regulation is a mediated process and as such, is subject to competition and saturation effects.
Mature small RNAs need to be loaded onto Argonautes in order to be activated.
This is a process which involves competition to access and be loaded to the Argonautes and can lead to saturation conditions for the underlying RISC-forming machinery of the cell.
This saturation can take place either through the large number of mature small RNAs present, or for example under Argonaute scarcity conditions.

The experimental evidence for competition and saturation effects in PTR comes from multiple sources. 
For example, it was shown that highly stable chemically modified siRNAs are able to outcompete the less stable native siRNAs\citep{Koller2006}.
Another study presented in vitro evidence of intra-siRNA competition and also siRNA-miRNA competition for Argonautes\citep{Castanotto2007}.
Additionally, up-regulation of target gene expression levels observed in transfection experiments was shown to be likely due to RISC saturation effects\citep{Khan2009}.
Regulation during the maturation process of small RNAs\citep{Kai2010}, miRNA sponges\citep{Ebert2010}, and RISC co-factors\citep{Hussein2011} 
can also induce competition-like and saturation-like effects of various strengths to different small RNAs.
On the other end of the competition spectrum, with Argonautes highly abundant, it has been found that small RNAs compete to access targets\citep{Loinger2012},
and that transcript abundance dilutes the small RNA activity\citep{Arvey2010}.
This is because transcripts with target sites for the same miRNA can compete for binding, in which case the titration of miRNAs by a certain target can release the repression
from other targets\citep{Pasquinelli2012}.

In the first part of this paper we construct a mathematical model of miRNA biogenesis and activity which explicitly takes into account RISC-formation.
Understanding how saturation and competition in RISC-formation can influence the dynamics of PTR is the second aspect of this work.
We will show that there is a distinctively different signature in the variation of target transcript expression depending on whether the mode of PTR is stoichiometric or catalytic.

\section*{Methods}

\textbf{Synthesizing a minimal model of PTR.}
We construct a model of post transcriptional regulation based on the loading of small RNAs onto Argonautes.
Our terminology is equally applicable to miRNA and siRNAs, and comparisons with experiments will draw on both prokaryotic and eukaryotic model systems.
The biogenesis of a single species of small RNA, the expression of targets, and RISC activity is modeled along the lines of previous work \citep{Levine2007,Hao2011}.
Creation and degradation of molecules of a particular small RNA species ($\bm m$), the target mRNA ($\bm R$), and the target protein ($\bm P$) are described by the following set of reactions
\begin{eqnarray}
&&\varnothing \begin{array}{c} \overset{k_m}{\rightarrow} \vspace*{-0.3cm} \\  \underset{d_m}{\leftarrow} \end{array} \bm{m},\;(\mathrm{small\;RNA})\label{r:m}\\
&&\varnothing \begin{array}{c} \overset{k_R}{\rightarrow} \vspace*{-0.3cm} \\ \underset{d_R}{\leftarrow} \end{array} \bm{R}, \;(\mathrm{target\;mRNA})\label{r:R}\\
&& \left. \!\!\!\begin{array}{l} \bm{R} \overset{k_P}{\rightarrow} \bm{R} + \bm{P} \\ \bm{P} \overset{d_P}{\rightarrow} \varnothing \end{array} \right\} 
\;(\mathrm{target\;Protein})\label{r:P}.
\end{eqnarray}
The values for all parameters are provided in Table~\ref{table:parameters}.
We denote the number of small RNA bound to Argonautes by $\bm A_m$, and the number of unbound Argonautes by $\bm A$. 
For simplicity we take the total number of Argonautes $A_0 = \bm A + \bm A_m$ to be constant. 
We then describe RISC assembly at rate $k_A$ and RISC disassociation at rate $d_A$ by
\begin{equation}
\bm{m} + \bm{A}  \begin{array}{c} \overset{k_A}{\rightarrow} \vspace*{-0.3cm} \\ \underset{d_A}{\leftarrow} \end{array} \bm{A_m} \;(\mathrm{RISC}).\label{r:Am}
\end{equation}
We stress that the RISC-formation process is independent of the RISC targeting activity, the latter described by the reaction 
\begin{equation}
\bm{A_m} + \bm{R} \overset{d_{Rm}}{\rightarrow} 
\left\{ \begin{array}{l} \bm{A} \mathrm{\;\;\;\;stoichiometric} \\ \bm{A_m} \mathrm{\;catalytic}\end{array} \right\}, (\mathrm{PTR})\label{r:Rm}
\end{equation}
with a rate $d_{Rm}$, and an outcome which depends on the mode of PTR operation, stoichiometric or catalytic.
For simplicity, we describe single-seed targets only, but our result are not conditional on this assumption,
since multi-seed targets effectively result in higher values of $d_{Rm}$\citep{Mukherji2011}.
According to  Eq.~(\ref{r:Rm}), a targeted mRNA via either mode of PTR cannot be released back into the pool of active mRNAs available for translation.
We consider the set of Eqs.~(\ref{r:m}-\ref{r:Rm}) as a minimal model of PTR mediated by small RNAs.

Under this minimal model, the number of Argonautes available is determined by the parameter $A_0$.
This parameter is affected by Argonaute biogenesis and turnover, as well as other small RNA species binding to Argonautes,
so it incorporates an element of the competition faced by the modeled small RNAs to access the Argonautes.
The actual formation of RISCs is determined by the rate $k_A$, which depends primarily on the individual efficiency that the small RNA species has to load to an Argonaute.
The abundance of available RISC loading factors, which in turn depends on the presence of other competing small RNA species, can influence $k_A$ as well.
Overall, the ratio $M_0 = d_A/k_A$ is a measure of the effectiveness of small RNAs at RISC-formation which also takes into consideration how stable the 
small RNAs are when incorporated in RISCs.
Low values of $M_0$ indicate that the small RNAs considered are highly effective at incorporating and stabilizing themselves in RISCs\citep{Haley2004}.
High values of $M_0$ reduce the inherent RISC-incorporating effectiveness, or increase the RISC-instability of the modeled small RNAs\citep{Castanotto2007}.
The presence of other equally potent or highly expressed small RNAs\citep{Koller2006}, or the saturation of co-factors involved in RISC-assembly\citep{Hussein2011}, 
are additional factors that contribute to high $M_0$ values.

In our study, the Fano factor of the number of molecules is the measure used to capture the strength of variation in different molecular species.
It is defined as the ratio of the variance over the mean of the expression level of a molecular species.
For example, a Fano factor of target transcripts equal to unity arises from mRNA expression and degradation at constant rates\citep{Thattai2001}.
A Fano factor exceeding unity indicates additional sources of mRNA variation~\citep{Paulsson2005}.
By harvesting repeated measurements under steady-state conditions, 
we collect sufficient statistics of the PTR module simulated using the Gillespie algorithm (direct method)\citep{Gillespie1977}.
In order to draw direct comparisons with previous theoretical work\citep{Hao2011}, the regime of PTR operation we present here is that of \textit{Salmonella}.
However, PTR in mammals was simulated as well, verifying that our results are not limited to PTR in bacteria only but extend over a broader scope.

\section*{Results}

\textbf{RISC-formation introduces an upper bound to the efficiency of small RNAs.}
The system of reactions described by Eqs.~(\ref{r:m}-\ref{r:Rm}) is approximated in the limit of large copy-numbers\citep{vanKampen}
by the following rate equations
\begin{align}
\frac{d\bm m}{dt} & = k_m - d_m\bm m - k_A\bm m\bm A + d_A\bm A_m, \label{d:m}\\
\frac{d\bm A_m}{dt} & = - \frac{d\bm A}{dt} = k_A\bm m\bm A - d_A\bm A_m \;\;\;\underbrace{-d_{Rm}\bm A_m\bm R}_{\textrm{stoichiometric}}, \label{d:Am}\\
\frac{d\bm R}{dt} & = k_R - d_R\bm R - d_{Rm}\bm A_m\bm R, \label{d:R}\\
\frac{d\bm P}{dt} & = k_P\bm R - d_P\bm P. \label{d:P}
\end{align}
The term over the bracket is present only in stoichiometric regulation.
If correlations among the constituents are neglected, an analytic solution for Eqs.~(\ref{d:m}-\ref{d:P}) can be found at steady-state.  
The part of the solution associated with RISC-formation, when combined with our assumption of $\bm A + \bm A_m = A_0$, yields
\begin{equation}
\bm A_m = A_0 \frac{\bm m/M_0}{1 + \bm m/M_0 \underbrace{+\bm R/R_0}_{\textrm{stoichiometric}}}. \label{ss:Am}
\end{equation}
For the constituents involved in RNA interference we find
\begin{gather}
\bm m = \frac{k_m}{d_m} \underbrace{-\frac{d_{Rm}}{d_m}\bm A_m \bm R}_{\textrm{stoichiometric}}, \label{ss:m}\\
\bm R = \frac{k_R/d_R}{1+ \frac{d_{Rm}}{d_R}\bm A_m}, \label{ss:R}
\end{gather}
where $\bm A_m$ is given by Eq.~(\ref{ss:Am}) and in stoichiometric PTR can depend also on the number of mRNAs.
In Eq.~(\ref{ss:Am}), $R_0 =  d_A/d_{Rm}$ is a threshold that emerges in the stoichiometric regime only and controls the targeting efficiency of the RISC complex before 
it is destabilized with the target.

We find a Michaelis-Menten functional form for Eq.~(\ref{ss:Am}), which is not surprising since the Argonautes operate in a manner similar to enzymes\citep{Haley2004}.
Experimental evidence for Eqs.~(\ref{ss:Am},\ref{ss:R}) comes from mammalian cell lines.
Cuccato et al.\citep{Cuccato2011} fit different models of PTR to data from cell lines expressing EGFP or tTA proteins.
Some of these models had a basis in explicit biochemical reactions, but the best fit was provided by a phenomenological model without a simple biochemical basis.
These best-fitting equations were of the form of Eqs.~(\ref{ss:Am},\ref{ss:R}) and we conclude that RISC-formation is the key mechanism behind their success. 

When the expression level of a particular small RNA increases in the
cell, this species is more likely to be loaded into Argonautes, be incorporated into RISCs and repress its targets.
In this case, the rate-limiting factor in the strength of target repression is set by $M_0$, i.e., the RISC-assembly mechanism.
When the small RNA expression reaches $\bm m \simeq M_0$, the target repression rate begins to level off as $A_m$ saturates due to competition between small
RNAs and saturation of the RISC-assembly machinery.
This effect is the basis behind particular experimental observations.
For example, it was found that differences of orders of magnitude among small RNAs and target transcripts do not suffice to fully suppress target expression\citep{Lim2005,Mukherji2011},
or that only moderate changes in transcriptome expression levels are observed when small RNAs are transfected into cells\citep{Baek2008,Selbach2008,Guo2010,Khan2009}.
It is not only the amount of small RNAs present in the cell relative to the number of targets that determines the strength of transcript suppression,
but also the relative competition that the small RNAs face in order to perform their tasks, and the number of targets that each small RNA has\citep{Arvey2010}.
The extra term in Eq.~(\ref{ss:Am}), which is present only in stoichiometric PTR, shows the mode of PTR operation plays also a crucial role in this context.
In stoichiometric PTR, RISCs are degraded with the targets, but in fully catalytic PTR this does not occur.
As a result, target derepression becomes easier under low RISC-recycling than high RISC-recycling conditions.

\textbf{RISC-assembly influences the variance of target transcript expression.}
Different species of small RNAs differ in their efficiency of RISC-incorporation.
In order to simplify the discussion, we present simulation data from two functional regimes of PTR operation:
the high efficiency (HE) regime characterized by $M_0 \ll \bm m$ values, 
and the low efficiency (LE) regime characterized by $M_0 \gg \bm m$ values.
We stress that the same small RNA species transfected or natively expressed in different cell types can find itself facing HE or LE conditions,
depending on the presence and expression levels of other small RNAs, and on the amount of freely available Argonaute proteins.
Furthermore, the mode of PTR can be affected by competition as well.
For example, it was found in mammalian cells that miRNAs are randomly sorted to the Argonautes, although the only Argonaute in mammals with RNA slicer activity is Ago2\citep{Wang2012}.
As a result, the same small RNA when loaded onto an Ago2-programmed RISC will perform cleaving-mediated PTR activity faster and possibly with higher RISC-recycling 
compared to Ago1,3,4-programmed RISCs, which rely on the action of nucleases and have in principle lower RISC-recycling.
The sorting of small RNAs to the Argonautes is random\citep{Wang2012}, therefore it is also highly susceptible to competition conditions induced by the expression of other small RNAs.

\textbf{Highly efficient small RNAs saturate the PTR pathway.}
First, we simulate cell transfections with highly efficient small RNAs and under conditions of Argonaute abundance.
Figure~\ref{fig:HE} shows the Fano factors and average molecular numbers per cell as functions of the target gene transcription rate $k_R$.
The case of HE stoichiometric PTR is shown in Figure~\ref{fig:HE}A.
In stoichiometric PTR, RISCs are degraded with the targets.
However, when the target transcription rate is lower than the small RNAs transcription rate ($k_R<k_m$), then
RISC-target degradation has a weak effect in RISC and substrate copy-numbers.
The highly efficient small RNAs titrate the Argonaute supply of the cell, as the inset of Figure~\ref{fig:HE}A shows, 
and additionally a number of free small RNAs remains as substrate.
On the other hand, targets with transcription rates close to the small RNA production rate ($k_R \simeq k_m$) are expressed at compensatory levels to those of the small RNAs,
and begin to alleviate the PTR-mediated repression\citep{Levine2007,Mukherji2011}.
This occurs via substrate depletion, since the newly available Argonautes from the degraded RISCs are quickly recruited by the highly efficient small RNA substrate.
The result is a very sharp transition from target repression to target derepression for those mRNAs with transcription rates $k_R \geq k_m$.
The main graph of Figure~\ref{fig:HE}A shows this sharp signature in expression level variation, especially for the substrate and RISCs.
Highly efficient small RNAs force RISC-target degradation to take place over a shorter $k_R$ range, and repression drops sharply for targets with $k_R > k_m$.
This property can be desirable for example for miRNAs or siRNAs involved in sharp activation and deactivation transitions of genes during the developmental phase of organisms.

Figure~\ref{fig:HE}B corresponds to HE catalytic PTR, where the range of the target transcription rate is increased 6-fold.
Here, there is no depletion of the free small RNA substrate taking place, and the RISC-target interaction does not degrade RISCs.
For this reason the repression-derepression threshold disappears, in agreement with experimental observations\citep{Mukherji2011}.
Titration of the Argonautes by the highly efficient small RNAs takes place again, as the inset of Figure~\ref{fig:HE}B shows, 
but this time over the full range of mRNA transcription rates.
The PTR module operates at maximum target repression capacity, and targets are strongly repressed and show small expression variation for any range of transcription rates.
Overall, we notice that targets with transcription rates far from the derepression threshold, whether catalytically or stoichiometrically regulated,
show variation of unity Fano factor strength.

Similar profiles in target expression variation were reported by a theoretical study which did not include the RISC-assembly process explicitly\citep{Hao2011}.
In view of our results and analysis so far, this is not surprising.
HE PTR conditions are conditions of fast RISC assembly, in which case an effective model which does not directly include this process is accurate.
However, it should be emphasized that HE conditions for a given small RNA gene, are LE conditions for the rest of the small RNA genes expressed in the cell.
If Argonautes can be as efficiently titrated by the product of a single or a few highly efficient small RNA genes as Figure~\ref{fig:HE} shows,
then the rest of the expressed small RNAs are facing Argonaute scarcity conditions.

\textbf{High RISC-recycling for less efficient small RNAs induces a novel signature in target expression variation.}
We now investigate post-transcriptional regulation mediated by small RNAs which have a low efficiency at RISC-formation. 
Figure~\ref{fig:LE}A shows the case of stoichiometric LE PTR.
As the free small RNA substrate is less responsive to RISC-formation, Argonaute titration becomes less efficient and fewer RISCs are formed,
the latter becoming easier to be degraded with targets.
The transition from target repression to target derepression becomes smoother, and the peak Fano factor value in target levels 
is reduced compared to the case of highly efficient small RNAs.
It should be mentioned that $d_{Rm}$, the rate at which RISCs target transcripts, is the same for both HE and LE conditions.
The term $d_{Rm}\bm A_m \bm R$ in Eq.~(\ref{d:R}) implies that target variation is affected by the RISC copy-number.
Fast RISC-formation leads to a larger pool of RISCs. Consequently, during RISC-depletion close to the derepression threshold, 
targets experience higher expression variation.
This is the reason why the peak of the Fano factor of target expression is lower for LE than HE small RNAs.
Overall, we find that in stoichiometric PTR the less efficient small RNAs, which are not demanding on PTR resources, can regulate targets 
as effectively as highly efficient small RNAs which tend to saturate the PTR network.
Viewed differently, our result suggests that stoichiometric PTR involving small RNAs of low efficiency has a capacity for many different 
independent ``channels", where the different types of small RNAs are concurrently involved in independent stoichiometric PTR tasks.

Figure~\ref{fig:LE}B shows the case of catalytic LE PTR with the $k_R$ range increased 6-fold.
As expected for PTR not operating at capacity, target derepression occurs at lower $k_R$ values compared to HE PTR.
However, a qualitatively different profile of PTR activity emerges.
The Fano factor, and hence the strength of variation in target mRNA numbers increases linearly with the target transcription rate.
Target transcripts with higher transcription rates than their regulators experience higher variation in their expression levels the higher $k_R$ becomes.
Target repression is again more effective compared to stoichiometric PTR,
but now the profiles in variation of target expression under stoichiometric and catalytic LE PTR are distinctively different.

It is known that the expression variation measured by the Fano factor for the biogenesis of mRNA under constant transcription and 
degradation rates is independent of the transcription rate~\citep{Thattai2001}.
This scenario is recovered far from the derepression threshold $k_R \gg k_m$ in stoichiometric LE PTR, and in both stoichiometric and catalytic HE PTR.
However, we find under catalytic LE PTR that the Fano factor becomes linearly dependent on the target transcription rate.
The root cause of this effect is suggested by Eq.~(\ref{ss:R}) showing that target abundance is affected by RISC abundance.
Due to the decreased ability  at RISC-formation of LE small RNAs, the average copy-number of targets responds to sustained changes in RISC numbers.
As a result, there are long time intervals where target turnover readjusts to changes in the number of RISCs.
Any increase in $k_R$ enlarges the pool of target transcripts $\bm R$, but also increases the turnover rate associated with the term $d_{Rm}\bm A_m\bm R$ of Eq.~(\ref{d:R}), 
simply because the same amount of RISC is now exposed to more targets.
Consequently, a linear increase of target expression variation with $k_R$ follows.
On the other hand, small RNAs efficient at RISC-formation induce fast changes to RISC copy-numbers, averaging over time to a constant rate of RNA interference activity.
This is the reason why the LE range of PTR operation could not be captured by previous models lacking the RISC-formation process\citep{Hao2011}.

\textbf{RISC-recycling determines the response of PTR to Argonaute scarcity conditions.}
We have discussed that Argonaute abundance conditions for small RNAs highly efficient in RISC-formation can result in Argonaute scarcity conditions for other 
less efficient but concurrently expressed small RNAs.
Additionally, highly efficient small RNAs due to competition from their isoforms, or from different small RNAs of similar efficiency, can face also Argonaute scarcity conditions.
So far, we have investigated cells with $A_0=500$ Argonautes available.
In what follows, we are going to probe lower values of $A_0$ for the LE operational regime of PTR.

Figure~\ref{fig:A0}A considers stoichiometric LE PTR close to the derepression threshold $k_R=k_m$.
As Argonaute availability increases, more small RNAs can be incorporated into RISCs, but on the other hand more RISCs can be degraded with targets.
Overall, the derepression threshold at $k_R=k_m$ remains fixed, but the average copy-number of substrate small RNAs and targets decreases with increasing $A_0$,
as Eqs.~(\ref{ss:m}-\ref{ss:R}) predict as well.
Consequently, as $A_0$ levels increase in the cell, $\bm m$ and $\bm A_m$ show greater expression level variation which affects also the expression level variation of targets.

The picture reverses for catalytic LE PTR, as shown in Figure~\ref{fig:A0}B.
RISCs can no longer be degraded with targets in catalytic PTR, which renders the number of RISCs independent of the number of targets.
On the other hand, the RISC copy-number is dependent on the Argonaute copy-number, 
but the amount of variation of the RISC copy-number is independent of $A_0$ as shown in Figure~\ref{fig:A0}B.
Furthermore, for the substrate small RNAs at steady-state, both the copy-number as Eq.~(\ref{ss:m}) predicts, and strength of variation become independent of $A_0$.
Lower Argonaute numbers reduce the PTR-mediated degradation rate of targets close to the basal degradation rate. 
Consequently, the lower $A_0$ becomes, the more that variation in RISC copy-numbers influences variation in the turnover of targets.
It should be mentioned that the monotonic increase of the Fano factors with decreasing $A_0$ shown in Figure~\ref{fig:A0}B is determined by the relative
strength of the PTR-mediated degradation rate of the target mRNA compared to the basal degradation rate.
At extreme Argonaute scarcity conditions, where $d_{Rm} \bm A_m \bm R \ll d_R$, a non-monotonic profile emerges, 
and an inflection point appears close to the crossover range $d_{Rm} \bm A_m \bm R \sim d_R$ (data not shown). 
In Figure~\ref{fig:A0}B, the crossover range corresponds essentially to $A_0=0$, therefore no inflection point is observed.

Under LE PTR, the impact of Argonaute scarcity conditions on target expression variation is dependent on RISC-recycling.
In the limit of weak RISC-recycling, a reduction in Argonaute availability reduces the strength of variation in target expression, 
but in the limit of strong RISC-recycling the opposite holds true.
This finding complements our conclusion from our study of stoichiometric regulation that less efficient small RNAs can be favored because they 
require less Argonautes to perform tasks equally well.
If multiple stoichiometric PTR ``channels" are open, the channel with the least efficient or lowest expressed small RNAs will be operating under Argonaute scarcity conditions.
However, as Figure~\ref{fig:A0}A shows, the level of noise in PTR for this channel will be reduced compared to the rest of the activated channels.
On the other hand, for catalytic regulation when multiple types of small RNAs are coexpressed, Argonaute scarcity conditions induce an increase to the strength of variation in the expression
of targets, leading to more noisy PTR activity.

\section*{Discussion}

Understanding competition effects within the PTR pathway is of central importance to PTR.
For example, transfection protocols using synthesized small RNAs to target particular genes,
require taking into account and optimizing for the strength of perturbation induced by the small RNA introduced to the cellular environment.
Highly expressed small RNAs, or small RNAs efficient in RISC-formation face low competition for RISC-loading in the transfected cell lines.
However, the natively expressed small RNAs in these cell lines face high competition, which may not be part of the physiological range of activity.
Endogenous aberrant expression of miRNAs in cancer cells can lead also to PTR competition conditions\citep{Esquela-kerscher2006}.
Here, abnormalities in the expression or efficiency of small RNAs due to mutations or DNA damage can perturb the rest of the normally functioning PTR pathway.

Accumulated insight into the mechanisms of PTR suggests that competition effects are ubiquitous also under normal physiological conditions.
For example, most mammalian miRNA genes have multiple isoforms\citep{Kim2009}.
When different isoforms have similar efficiency at RISC-formation, one can expect competition to set in even when low numbers of miRNA or siRNA genes are expressed.
Similar conditions also arise for PTR in prokaryotes, with differences to eukaryotic PTR being quantitative rather than qualitative\citep{Vanderoost2009}.

It is important to understand that within the same cellular environment, in principle different PTR conditions can apply for the different small RNA species and their targets.
If multiple highly efficient or abundant small RNA genes are natively expressed, their targets will exhibit the variation predicted under LE instead of HE PTR conditions, 
since the different species of small RNAs will compete for the Argonautes.
Additionally, targets of the rest of the natively expressed small RNAs in these cell lines might experience PTR under Argonaute scarcity conditions,
if the majority of the Argonautes are titrated by the more efficient small RNAs.

We have developed a mathematical model of RISC-formation from which several principles of post-transcriptional regulation emerge.
Argonaute proteins operate on the substrate of free small RNAs as enzymes, and when modeled as such, the functional form of Eqs.~(\ref{ss:Am},\ref{ss:R}) emerges.
This functional form was used as a heuristic without biochemical basis to model PTR data \citep{Cuccato2011}, where it gave a superior fit to the data
compared with several other models not incorporating RISC-formation. 
Our results indicate that the upper bound in the efficiency of small RNAs is set not only by the number of Argonautes in the cell but also by the 
RISC-degradation rate $d_A$ present in the RISC-saturation thresholds $M_0$ and $R_0$ in Eq.~(\ref{ss:Am}).
Furthermore, RISC turnover is affected in stoichiometric PTR by the joint RISC-target degradation process and becomes dependent on $d_{Rm}$ as well.

Small RNAs play a crucial role during the developmental phase of organisms by activating or deactivating networks of genes, or by maintaining expression thresholds.
Our mathematical model predicts that in stoichiometric PTR the target derepression transition is sharp under HE conditions but smoother and less noisy under LE conditions.
On the one hand, highly efficient small RNAs can perform their tasks better and faster.
On the other hand, they can outcompete the less efficient small RNAs concurrently involved in PTR, thus can saturate the capacity of the PTR pathway for multi-tasking.
The profiles of stoichiometric PTR under LE and HE conditions are similar, but small RNAs less efficient at RISC-formation use fewer of shared cellular resources.
This opens up the possibility for greater numbers of independent PTR pathways to coexist, allowing for more complex cellular functionality to emerge.

Mukherji et al.\citep{Mukherji2011} reported a similar repression-derepression pattern when they administered miRNA-mimicking siRNAs to aid miR-20 in target repression,
or introduced additional target seed sites.
Both measures did not change the RISC-loading efficiency of miR-20, but led to an increase in the target repression efficiency via 
an effective increase in the value of $d_{Rm}$. 
This increase of repression efficiency resulted in a sharpening of the repression-derepression threshold.
Our analysis shows that even when the targeting efficiency of miRNAs is held fixed by keeping $d_{Rm}$ constant, one would still expect differences in the sharpness of 
the derepression threshold due to differences in the RISC-loading efficiency of the miRNA species.

Small RNAs with low RISC-formation efficiency induce a novel signature in the variation of target expression levels.
Although target suppression is less severe compared to PTR mediated by highly efficient small RNAs, our model predicts that the variation in the target expression levels 
(measured by the Fano factor) increases linearly with the target transcription rate.
Transcripts targeted catalytically and kept at a threshold level show larger expression variation around the threshold when they are targeted
under low efficiency than high efficiency PTR conditions.
This finding suggests that for example an organism with two post-transcriptionally regulated paralogous genes, one with accumulated mutations in the seed-region,
and the other perfectly complementary to the small RNA regulators, will show different RNA interference dynamics, depending on which copy is expressed.
The transcripts with imperfect complementarity to the small RNA
regulators tend to follow the stoichiometric mode of PTR and will be stably repressed or kept at a threshold.
The transcripts with perfect complementarity follow the catalytic mode and when kept at a threshold will show greater expression level variance.
Essentially, this organism will possess the capacity to control the expression profile of a particular protein simply by activating the corresponding copy of the paralogous gene.

We have presented data of PTR corresponding to two limiting RISC-recycling rates: stoichiometric and catalytic.
However, our investigation was extended also to intermediate RISC-recycling rates.
Highly efficient small RNAs of intermediate RISC-recycling were studied before\citep{Hao2011}.
For low efficiency small RNAs we find that the fully catalytic profile of Figure~\ref{fig:LE}B is modified and an inflection point 
appears at a level of target expression where the joint RISC-target degradation is able to deplete the small RNA substrate (data not shown).
The higher the RISC-recycling rate, the further away in $k_R\gg k_m$ values this inflection point appears.
Overall, our results and conclusions based on stoichiometric and catalytic regulation are not significantly modified.
High RISC-recycling rates tend to reproduce the fully catalytic PTR profile, and low RISC-recycling rates approach the stoichiometric PTR profile.

Recent experiments identified the average number of Argonautes in mammalian cells to be of the order of $10^5$ proteins per cell\citep{Wang2012}.
In human 293T cells, it is found that unregulated targets of 15000-25000 copies per cell are reduced approximately 5-fold by miR-223 reaching levels of 2700 copies per cell\citep{Baccarini2011}.
In order to test if our results and conclusions based on PTR in bacteria apply to other organisms, 
we have simulated PTR in mammals using the above expression levels (data not shown). 
We find qualitatively similar results and we conclude that our predictions hold also for mammalian PTR.

In summary, our work elucidates how the molecular machinery behind PTR, and in particular competition of miRNAs and siRNAs during RISC-formation
can influence the dynamics in target expression. 
We find two distinct regimes of PTR activity associated with low and high variation in the levels of target transcripts.
It is possible for future work to construct a protocol in order to infer the mode of PTR, stoichiometric or catalytic, based on the variation of target
expression levels, as well as elucidate the effects of Argonaute competition across different species of small RNA.

\section*{Acknowledgments}
This work was supported by Deutsche Forschungsgemeinschaft (DFG) grant
SFB 680, by the BMBF-SysMO2 grant and by SyBaCol. We acknowledge
discussions with Zuzanna Makowska, Marvin Jens, Alexander Loewer, and Nikolaus Rajewsky.

\bibliography{references}

\clearpage

\section*{Tables}

\begin{table}[!ht]
\begin{center}
\begin{tabular}{|c|c|c|}
\hline
\footnotesize parameter & \footnotesize definition & \footnotesize value \\\hline\hline
$\bm m$ & \footnotesize small RNA number & \footnotesize - \\\hline
$\bm R$ & \footnotesize target mRNA number & \footnotesize - \\\hline
$\bm P$ & \footnotesize target protein number & \footnotesize - \\\hline\hline
$k_{m}$ & \footnotesize small RNA production rate & \footnotesize 1/min \\\hline
$d_{m}$ & \footnotesize small RNA degradation rate & \footnotesize 1/h \\\hline\hline
$k_{R}$ & \footnotesize mRNA transcription rate & \footnotesize 1/min \\\hline
$d_{R}$ & \footnotesize mRNA basal degradation rate & \footnotesize 1/h \\\hline
$d_{Rm}$ & \footnotesize target destabilization rate & \footnotesize 1/h \\\hline
$k_{P}$ & \footnotesize mRNA translation rate & \footnotesize 2/h \\\hline
$d_{P}$ & \footnotesize protein degradation rate & \footnotesize 0.5/h \\\hline\hline
$k_{A}$ & \footnotesize RISC assembly rate & \footnotesize (see $M_0$) \\\hline
$d_{A}$ & \footnotesize RISC disassembly rate & \footnotesize (see $M_0$) \\\hline\hline
$A_0$ & \footnotesize Argonaute number & \footnotesize 500 \\\hline
$M_0$ & \footnotesize $d_A/k_A$ & \footnotesize 1/600 - 600 \\\hline
$R_0$ & \footnotesize $d_A/d_{Rm}$ & \footnotesize 0.1 - 60 \\\hline
\end{tabular}
\caption{
{\bf Definitions of model parameters.}
Typical values correspond to the activity of siRNAs in \textit{Salmonella}\citep{Overgaard2009,Hao2011}.
Rates are expressed in cell volume units.
\label{table:parameters}
}
\end{center}
\end{table}

\clearpage

\section*{Figure Legends}

\begin{figure}[!htb]
\begin{center}
\end{center}
\caption{\small
{\bf PTR under high efficiency of RISC-formation.}
Fano factors (main graphs) and average molecular numbers per cell (insets) of the free small RNAs $\bm m$ (yellow), loaded RISCs $\bm A_m$ (blue) 
and free target transcript $\bm R$ (red) are plotted as functions of the target gene transcription rate $k_R$ under low competition PTR conditions ($M_0=1/600$).
Colored axis lines and axis labels indicate the corresponding quantity they refer to.
Small RNAs have a fixed rate of production $k_m$=1/min which determines the derepression threshold\citep{Levine2007}.
{\bf (A)} Stoichiometric PTR. RISC and substrate degradation occurs in a short range around $k_R=k_m$, and the transition from repressed to derepressed targets is sharp.
{\bf (B)} Catalytic PTR at 6-fold increased $k_R$ range. 
Argonautes are titrated by the highly efficient substrate throughout the $k_R$ range because RISCs are not degraded with the targets, 
which leads to strong target repression.
}
\label{fig:HE}
\end{figure}

\begin{figure}[!htb]
\begin{center}
\end{center}
\caption{\small
{\bf PTR under low efficiency of RISC-formation.}
Similar to Figure~\ref{fig:HE} but for low efficiency small RNAs ($M_0=600$).
{\bf (A)} Stoichiometric PTR. 
Less efficient small RNAs are incorporated to lower numbers of Argonautes.
The target repression to target derepression transition becomes smoother.
{\bf (B)} Catalytic PTR at 6-fold increased $k_R$ range.  
A distinctively different signature in target expression level variation is found. 
The strength in target repression is reduced compared to HE catalytic PTR, but the level of variation in target expression increases linearly with $k_R$.
}
\label{fig:LE}
\end{figure}

\begin{figure}[!htb]
\begin{center}
\end{center}
\caption{\small
{\bf RISC-recycling determines the response of PTR to Argonaute scarcity conditions.}
The Fano factors of substrate small RNAs (yellow), loaded RISCs (blue), target transcripts (red) and target protein (gray) are plotted as functions of the Argonaute 
abundance $A_0$ for LE small RNAs.
{\bf (A)} Stoichiometric PTR near the derepression threshold $k_R=k_m=1/$min. 
More RISCs are formed as more Argonautes become available.
However, RISC turnover increases also due to the joint RISC-target degradation, which results in an increase in the strength of target expression variation.
{\bf (B)} Catalytic PTR at $k_R=12k_m$. 
Absence of joint RISC-target degradation renders the copy-number of small RNAs and the strength of variation of RISC copy-number independent of $A_0$.
Lower numbers of available Argonautes lead to a reduced PTR-mediated target degradation rate, which approaches the basal target degradation rate, 
resulting in increased levels of variation in target copy-numbers.
}
\label{fig:A0}
\end{figure}

\clearpage

\begin{figure}[!htb]
\begin{center}
\includegraphics[totalheight=16.0cm, keepaspectratio=true, viewport=0 0 785 1450,clip]{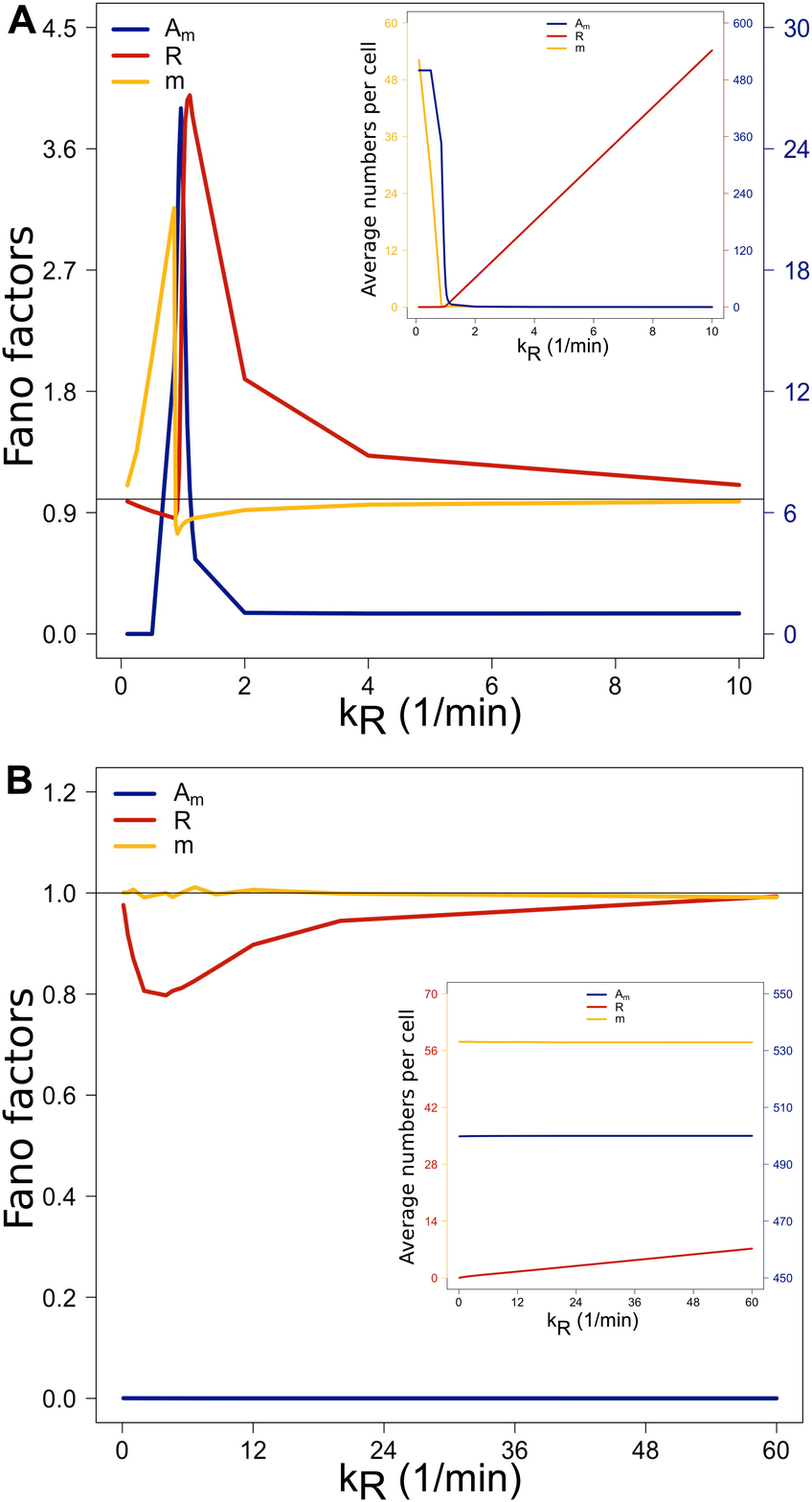}
\end{center}
\end{figure}

\begin{center}
Figure 1
\end{center}

\clearpage

\begin{figure}[!htb]
\begin{center}
\includegraphics[totalheight=16.0cm, keepaspectratio=true, viewport=0 0 790 1450,clip]{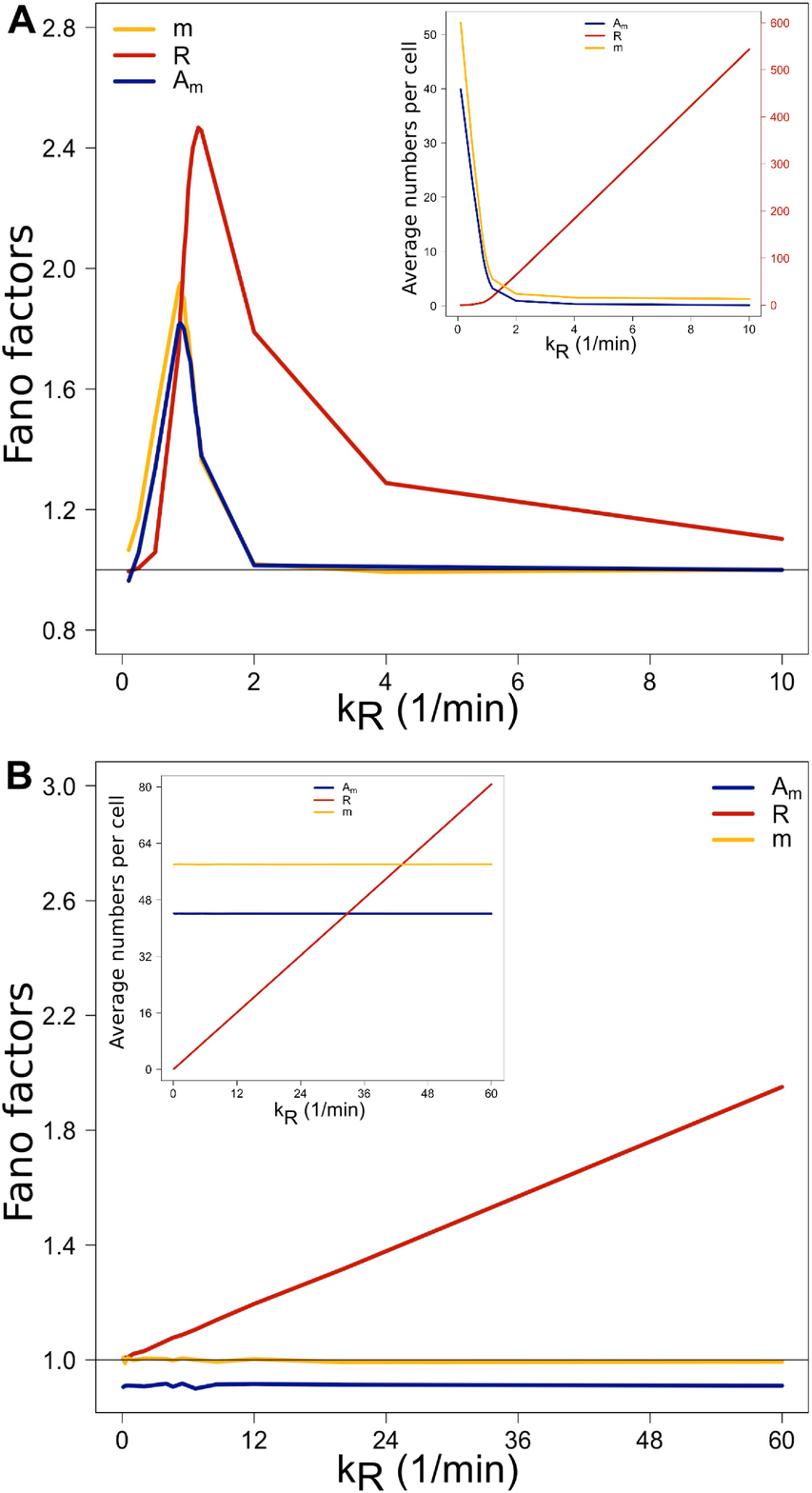}
\end{center}
\end{figure}

\begin{center}
Figure 2
\end{center}

\clearpage

\begin{figure}[!htb]
\begin{center}
\includegraphics[totalheight=16.0cm, keepaspectratio=true, viewport=0 0 800 1440,clip]{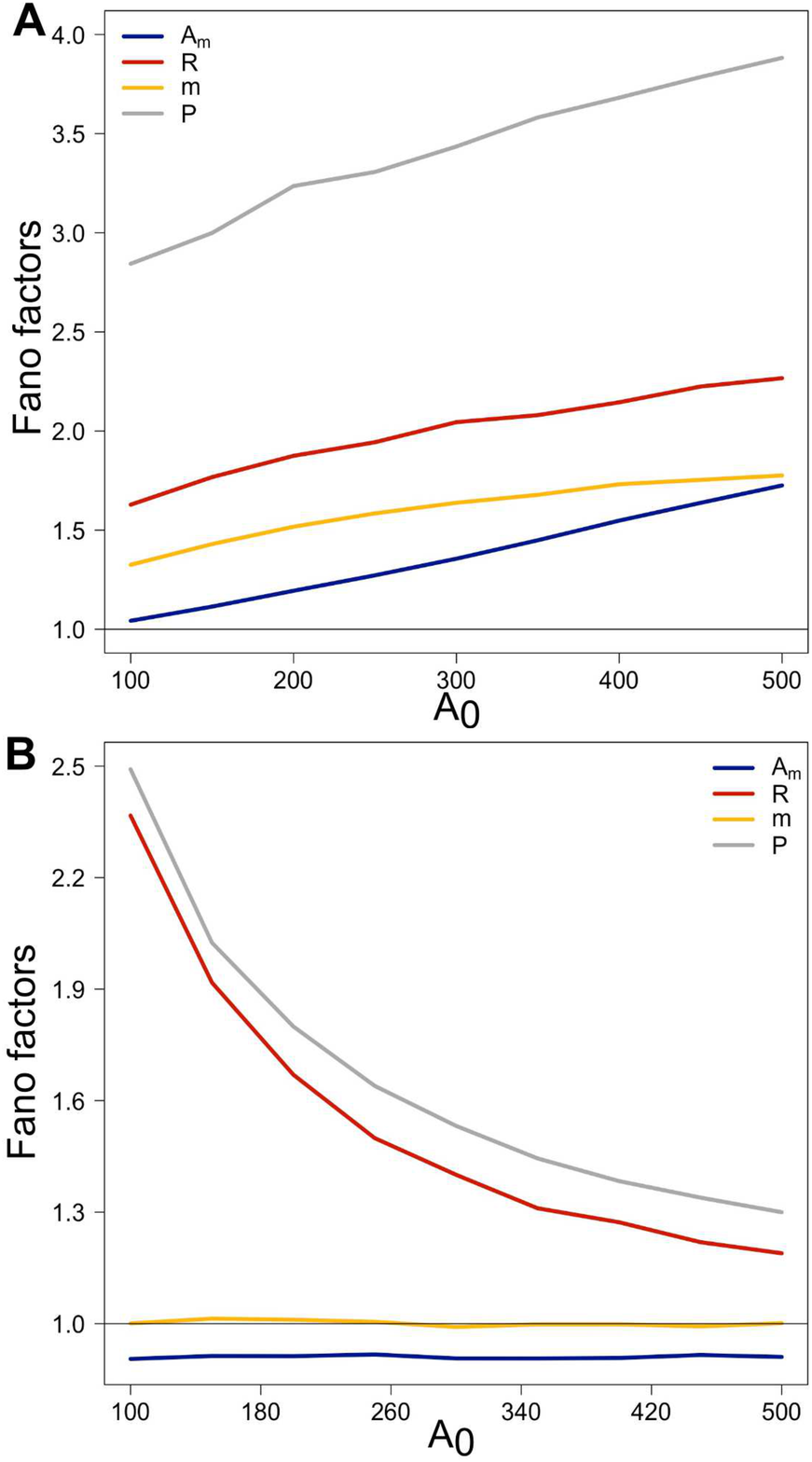}
\end{center}
\end{figure}

\begin{center}
Figure 3
\end{center}



\end{document}